\pdfoutput=1

\documentclass[letterpaper,twocolumn,10pt]{article}
\usepackage{template}

\usepackage{tikz}
\usepackage{amsmath}
\usepackage{color,xcolor}
\usepackage{filecontents}
\usepackage{verbatim}
\usepackage{multirow}
\usepackage{booktabs}
\usepackage{pgfplots}
\pgfplotsset{compat=1.17}

\begin{document}

\date{}

\title{\Large \bf PersonaMark: Personalized LLM watermarking for model protection and user attribution\\
}

\author{
{\rm Yuehan Zhang}\\
Wuhan University
\and
{\rm Peizhuo Lv}\\
Chinese Academy of Sciences
\and
{\rm Yinpeng Liu}\\
Wuhan University
\and
{\rm Yongqiang Ma}\\
Wuhan University
\and
{\rm Wei Lu}\\
Wuhan University
\and
{\rm Xiaofeng Wang}\\
Indiana University Bloomington
\and
{\rm Xiaozhong Liu}\\
Worcester Polytechnic Institute
\and
{\rm Jiawei Liu}\thanks{Corresponding Author.}\\
Wuhan University
} 

\maketitle

\begin{abstract}

The rapid advancement of customized Large Language Models (LLMs) offers considerable convenience. However, it also intensifies concerns regarding the protection of copyright/confidential information. With the extensive adoption of private LLMs, safeguarding model copyright and ensuring data privacy have become critical. Text watermarking has emerged as a viable solution for detecting AI-generated content and protecting models. However, existing methods fall short in providing individualized watermarks for each user, a critical feature for enhancing accountability and traceability. In this paper, we introduce PersonaMark, a novel personalized text watermarking scheme designed to protect LLMs' copyrights and bolstering accountability. PersonaMark leverages sentence structure as a subtle carrier of watermark information and optimizes the generation process to maintain the natural output of the model. By employing a personalized hashing function, unique watermarks are embedded for each user, enabling high-quality text generation without compromising the model's performance. This approach is both time-efficient and scalable, capable of handling large numbers of users through a multi-user hashing mechanism. To the best of our knowledge, this is a pioneer study to explore personalized watermarking in LLMs. We conduct extensive evaluations across four LLMs, analyzing various metrics such as perplexity, sentiment, alignment, and readability. The results validate that PersonaMark preserves text quality, ensures unbiased watermark insertion, and offers robust watermark detection capabilities, all while maintaining the model's behavior with minimal disruption.

\end{abstract}

\section{Introduction}

Since the inception of ChatGPT, Large Language Models (LLMs) have garnered significant attention, spearheading advancements across various domains, including automatic news composition, social media content generation, and creative writing assistance. These models have become essential tools, impacting numerous facets of daily life \cite{chen_soulchat_2023,huang_lawyer_2023,dan_educhat_2023,luo_yayi_2023}.  In parallel, the growth of customized and proprietary LLMs tailored to specific fields has introduced new ethical, legal, and social challenges. These challenges encompass concerns surrounding content authenticity, copyright, intellectual property (IP) protection, and the regulation of AI-generated content. Organizations developing domain-specific LLMs encounter substantial hurdles in protecting proprietary knowledge and copyrighted material.

Among the most susceptible are personalized Large Language Models (LLMs), which necessitate robust copyright protections. These models are designed to emulate specific tones, personalities, or speech patterns—often aligned with the Myers-Briggs Type Indicator (MBTI) or other personal characteristics—and are increasingly utilized in various professional and personal contexts \cite{teper_leveraging_2024,wozniak_personalized_2024,neelakanteswara_rags_2024,noauthor_characterai_nodate,sabour2022chatbotsmentalhealthsupport}. As the trend towards personalized LLMs grows, ensuring the authenticity and control of their generated outputs becomes crucial.
Examples of such models include Character.AI \cite{noauthor_characterai_nodate}, CharacterGLM\cite{zhou_characterglm_2023}, and ChatPLUG\cite{tian2023chatplugopendomaingenerativedialogue}, which consistently produce outputs that reflect distinct personalities and stable character traits, including unique language styles. 

\begin{figure*}[!t]
    \centering
    \includegraphics[width=1\linewidth]{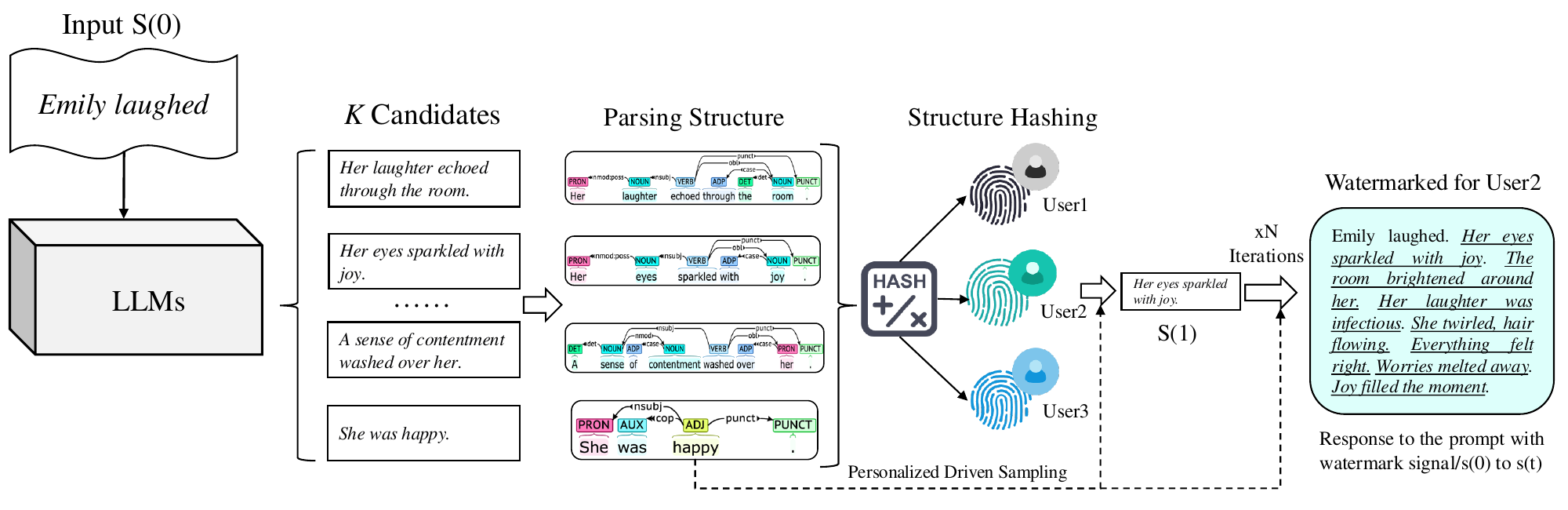}
    \caption{Overview of our proposed PersonaMark watermark injection framework. The \textit{italicized} and underlined sentences are watermarked sentences.}
    \label{fig:overview_mark}
\end{figure*}

Customized models, when commercialized and made accessible to the public, present significant concerns regarding potential misuse. For example, a personalized LLM trained to act as a famous host could provide users with interesting interactive experience, but it can also be used to campaign for someone, such as helping Trump win votes. Additionally, these models could potentially be misused to generate pornographic content, which is unacceptable. Whether pertaining to hosts, artists, or politicians, there is a collective intention to prevent models trained with personal data from being used to produce misinformation and mislead the public. Furthermore, there exists a risk of copyrighted information being misappropriated and misused. For example, a pharmaceutical company specializing in drugs for mental illnesses may use proprietary data to train drug discovery models for research and development purposes. However, these models might also be co-opted for the illegal production of substances, such as novel psychoactive drugs \cite{azai_ai_drug_2023,skinnider_deep_novel_psy_2021,noauthor_new_nodate_criminals}. In scenarios of leakage and misuse, it is imperative that companies possess the capability to trace such activities to specific users or employees and hold individuals accountable.

As traditional post-hoc detection methods falter, text watermarking has emerged as a promising solution to protect LLM-generated content \cite{zhang2024detection,yang2023watermarking,hou2023semstamp}. Recent advances in text watermarking embed imperceptible signals, or watermark messages, into text data, making them detectable by algorithms but invisible to human readers\cite{kirchenbauer_watermark_2023,zhao_provable_2023,liu_unforgeable_2023,lee_who_2023,piet_mark_2023,zhang_remark-llm_2023,tang_did_2023,hu_unbiased_2023,wu_dipmark_2023,yoo_advancing_2023,kirchenbauer_reliability_2023,kirchenbauer2024reliabilitywatermarkslargelanguage,liu_semantic_2023,hou2023semstamp,munyer_deeptextmark_2024}. These signals are embedded either during text generation or through post-processing.  Current techniques often manipulate token distributions by dividing vocabulary into ``green'' and ``red'' lists, promoting certain tokens while suppressing others.

However, these methods lack the robust protection and tracing capabilities characteristic of traditional image watermarking. Personalized watermarking, which involves embedding distinct watermarks into content for different users, is crucial for copyright protection in customized LLMs. Yet, current works are insufficient to meet this needs. They predominantly focus on detecting AI-generated text and rely on embedding simple binary watermark signals, failing to produce unique watermarks for individual users. Even multi-bit watermarking techniques suffer from limited payload capacity\cite{liu2023survey}, which compromises both robustness and text quality, thereby restricting their ability to manage extensive user bases\cite{yoo-etal-2024-advancing} and to conduct effective tracing. Moreover, the manipulation of token probabilities to inject watermarks introduces biases that conflict with the training objectives of LLMs, ultimately degrading the quality of the generated text. Consequently, existing methods are inadequate for achieving the same level of protection and traceability as traditional watermarking techniques, highlighting the need for more advanced approaches in the context of personalized LLMs.

To address the challenges mentioned above, we propose a novel method personalized text watermarking based on sentence structure, termed \textbf{PersonaMark}. As illustrated in Figure \ref{fig:overview_mark}, PersonaMark is a plug-in framework that enables LLMs to embed unique, user-specific watermarks into generated text while maintaining high content quality and preserving model behavior. The core motivation behind PersonaMark is to overcome the limitations of existing text watermarking techniques, which frequently fall short in providing robust user-specific watermarking without compromising text quality. 

PersonaMark operates by embedding personalized watermarks into text based on the structure of sentences. The process begins with the generation of $N$ candidate sentences for each prompt using a diverse beam search. The structure of each candidate sentence is extracted using a dependency parser and is converted into a string representation. These structures are then input into a hashing function, which partitions the sentence structure space into two equal parts, assigning a binary output (0 or 1) to each. 
Only sentences corresponding to a hash output of 1 are retained, with the top-ranked sentence (based on beam score) selected as the final output for the current step. This process repeats for each subsequent sentence until the entire prompt is completed. The result is a paragraph where all sentences share the ``1'' feature, which can be reliably detected through a statistical $Z$-test. The $Z$-test determines whether the null hypothesis ($H_0$) that the numbers of 0 and 1 outputs are equal, should be rejected. If rejected, it indicates the presence of the watermark.

To enable user-specific watermarking, the hashing function is designed to behave uniquely for each user, based on their user ID. This customization creates a distinct dividing plane in the sentence structure space for each user, ensuring that the watermarking process is individualized. During detection, the dividing plane corresponding to the detected text is identified, which allows tracing the content back to the specific user ID. This approach is notably scalable, capable of managing a large number of users while maintaining the quality and integrity of the generated text. 
By focusing on sentence structure rather than token manipulation, PersonaMark maintains the naturalness of the text while providing a robust mechanism for watermarking and attribution, addressing the critical need for copyright protection in the realm of personalized LLMs.

We conducted experiments across four model series using three prompt datasets: story writing, fake news generation, and report generation. Our method significantly outperformed existing approaches, achieving a perplexity of 2.74 compared to the baseline KGW's 12.32 \cite{kirchenbauer_watermark_2023}. Furthermore, our approach closely matched the un-watermarked model in terms of sentiment values, key information similarity, and sentence readability scores, perserving the original model behaviors. In terms of watermark robustness, our method excelled against various attack granularities, demonstrating its effectiveness for copyright protection and user attribution. Notably, under 40\% rate of word-level substitution attack, the AUC score of watermark detection remains 0.97. PersonaMark provides a reliable means to control the use of private LLMs and accurately identify specific users in cases of misuse or illegal activity.

\textbf{Contributions:} We summarize our contributions as below:

(1) We propose the personalized text watermarking approach that embeds specific watermarks based on unique user IDs. This method is highly scalable, making it suitable for protecting customized and stylized LLMs with a large amount of users with fast attribution speed.

(2) We utilize the sentence structures and user-specific hashing to achieve sentence-level watermarking. The watermark is intricately linked to language style, making it difficult to remove without altering inherent styles. Results indicate that text quality is well-preserved with minimal impact on model behaviors, thereby enhancing the approach's viability for real-world applications.

(3) We introduce three indicators designed to comprehensively assess the impact of watermarking techniques on model behavior and text outputs. Extensive experiments results demonstrate the superiority of our proposed method in terms of perplexity, sentiment polarity, key information consistency, readability and detection accuracy.

\section{Background and Related Work}

\subsection{Personalized LLM}

As the result of previous mentioned trend and motivations, there emerges many scenarios of personalized LLMs such as role playing, LLM personalities and value alignment for individual users. For role playing, the models may acquire knowledge and language styles of specific roles such as Harry Potter or roles in Haruhi Suzumiya. The role playing models are trained to act as one or more roles\cite{yu2024neekoleveragingdynamiclora, cui2023thespianmulticharactertextroleplaying}. For LLM personalities, multimodal models may have deeply personalized character and act like the person in the TV show by datasets constructed from famous TV series such as The Big Bang Theory\cite{xuanyuan2023hisheldoncreatingdeep}. The users will feel that they are chatting with the exact role in the novels, TV shows or even with the famous persons in the real world. 
There are not only personalized LLMs train by one person's character, but also personalized LLMs trained to have general personalities. These models are often trained or fine-tuned to have and act out the common features of personalities, which are often classified or divided by Myers–Briggs Type Indicator (MBTI)\cite{rao-etal-2023-chatgpt}, Big Five Factors\cite{10.5555/3666122.3666588}.
As personal LLMs gain popularity, users might seek AI companions like a virtual girlfriend or close friend. Additionally, individuals could train a digital "twin" to handle tasks and responses during their absence, creating a more personalized and interactive experience.\cite{noauthor_people_2023}. This can be easily achieved by techniques such as supervised fine-tuning, RLHF, and prompting.

\subsection{Sentence Structure and Parsing}

A sentence has many aspects, including syntactic ones and synthetic ones. The syntactic features which often appear in the form of sentence structures could be used to represent a sentence. Syntactic analysis, also known as parsing, is one of the core components in the field of natural language processing (NLP). It is a process of structured analysis of natural language texts and is of great significance for language understanding. This process aims to reveal the grammatical roles and interrelationships of vocabulary in text, enabling computers to understand and interpret the complexity of human language. Through parsing, natural language could be converted into a form that computers can process, which is crucial for applications such as machine translation, sentiment analysis, and information extraction. The parsing process sometimes checks the semantic rationality of the text based on the rule system of formal grammar.

During the parsing process, algorithms recognize vocabulary in the text and determine how these words are combined into phrases, sentences, and other structures based on preset grammar rules. This usually involves identifying parts of speech such as nouns, verbs, adjectives, and their functions in the sentence, such as subject, predicate, object, etc. The results of parsing are usually presented in the form of a tree structure, which clearly displays the hierarchical and dependency relationships between vocabulary.

Syntax parsing focuses on analyzing the grammatical structure of sentences, with core tasks including part of speech tagging, recognition of syntactic structures, and syntactic dependency relationships between vocabulary. Within the scope of syntactic parsing, Constituency Parsing reveals the hierarchical structure of a sentence by constructing Parse Trees, while Dependency Parsing focuses on revealing the dependency relationships between words, which are typically presented in a tree like structure to reflect the grammatical connections between words.

The key elements of dependency grammar include: 1. Dependency Relations. Within the framework of dependency grammar, each lexical element in a sentence has a direct dependency relationship with at least one other lexical element. These relationships can manifest as dominant (such as the dominance of verbs over objects) or modifying (such as the modification of nouns by adjectives). Dependency relationships are usually represented by arrow symbols, with arrows pointing to the head and from the head to the dependent. 2. Dependency Tree. The output of dependency parsing is usually presented in the form of a dependency tree, which is a type of Directed Acyclic Graph (DAG) where nodes represent lexical elements and edges represent dependency relationships. The root node of a dependency tree is usually the subject or predicate of a sentence, while other nodes are organized based on their dependency relationship with the root node. 3. Dependency Labeling. In dependency grammar, dependency relationships are typically assigned specific labels to indicate the type of relationship, such as "nsubj" (subject dependency), "obj" (object dependency), "amod" (adjective modifying dependency), etc.

The other two theories related to dependency syntax are Case Theory and Representation Theory, which assist in understanding language structure and function. The theory of case focuses on the internal construction of noun phrases (NP), particularly the interaction between nouns and their modifying components. Under this theoretical framework, nouns reflect their grammatical functions and semantic roles in syntactic structures through different cases, such as nominative, objective, and possessive. The core goal of case theory is to elucidate the internal organizational principles of noun phrases and how nouns achieve semantic diversity through changes in case. In dependency syntax analysis, the dependency terms of nouns may include modifiers of their case markers, and case theory explains how these modifiers interact with nouns and their semantic functions in sentences.

\subsection{LLM watermarking}

Kirchenbauer et al.\cite{kirchenbauer_watermark_2023} proposed a digital watermarking framework for large models that can embed signals that are invisible to humans but can be detected by algorithms in the generated text to mitigate the potential harm of large models. This framework(referred to as KGW in the following sections) achieves watermark embedding and detection by selecting a set of "green" markers and prioritizing their use during the generation process. Lee et al.\cite{lee_who_2023} proposed a code generation watermarking method called SWEET, which improves the ability to detect AI generated code text by removing low entropy segments during watermark generation and detection, while significantly improving code quality preservation, surpassing all baseline methods, including post detection methods. Hu et al.\cite{hu_unbiased_2023} proposed an unbiased watermarking technique that enables the tracking of model generated text without affecting the output quality of large models (LLMs), while ensuring that users cannot detect the presence of watermarks, providing a new perspective for responsible AI development.

DiPmark\cite{wu_dipmark_2023} protects data by embedding hidden information in the text generated by large models, while maintaining the original data distribution unchanged, achieving concealment, efficiency, and elasticity, making it a powerful solution for text attribution and quality protection. Yoo et al.\cite{yoo_advancing_2023} proposed a method called "Multi bit Watermark Allocation by Position" (MPAC), which embeds traceable multi bit information in the process of generating large models to address the problem of abuse in AI generated text. It effectively embeds and extracts long messages (32 bits). Although it holds potential applicability in user attribution scenarios, its detection accuracy diminishes substantially with an increase in bit number, as the vocabulary is partitioned into more segments rather than the original two-part division employed by KGW\cite{kirchenbauer_watermark_2023}. User attribution necessitates the management of a substantial number of accounts while ensuring that robustness and detection accuracy remain uncompromised as the user base expands. These are precisely the features of our method, as our initial design was specifically tailored for multi-user environments. Wang et al.\cite{wang_towards_2023} proposed a novel encoding text watermarking method called Balance Marking, which is used to embed multi bit information in text generated by large models (LLMs). This method balances the probabilities of available and unavailable vocabulary by introducing a surrogate language model, effectively improving the quality and diversity of generated text without significantly increasing computational complexity. Kirchenbauer et al.\cite{kirchenbauer_reliability_2023} explored the reliability of watermarking technology in detecting and recording text generated by Large Models (LLMs) in real-world scenarios through empirical analysis. They found that watermarks can still be detected even when humans and machines rewrite the text, indicating that watermarking technology has high robustness in resisting various attacks and modifications. Liu et al.\cite{liu_semantic_2023} proposed a semantically invariant robust watermarking method that utilizes semantic embeddings generated by auxiliary language models to determine watermark logic, achieving high-precision detection of large model generated text while achieving a good balance between attack robustness and security robustness.

Christ et al.\cite{christ_undetectable_nodate} proposed an undetectable watermark scheme based on cryptographic principles for language models, ensuring that only users with secret keys can detect the watermark without changing the output distribution, thereby achieving covert labeling of AI generated text while maintaining text quality. Kuditipudi et al.\cite{kuditipudi_robust_2023} proposed a novel robust watermarking method that does not alter the distribution of text. By mapping a random number sequence to text samples generated by a language model, reliable labeling of the text source is achieved, and the watermark detection ability is maintained even after the text has been edited or cropped. Hou et al.\cite{hou_semstamp_2023} proposed SEMSTAMP, a sentence level semantic watermarking algorithm based on local sensitive hashing (LSH). By encoding and partitioning the semantic space in the sentence embedding space, it enhances the robustness against rewriting attacks while improving the detection ability of AI generated text while maintaining generation quality.

The existing research on watermark algorithms in the generation process has insufficient attention to the methods of adding watermark information at the sentence level and the token sampling stage, and has introduced much bias in the watermark addition process. The existing watermarking algorithms mainly add watermark information at the word level or token level, and pay little attention on sentence level; At the same time, most existing research is limited to modifying logits probabilities and neglects the sampling stage. Modifying during the logits generation stage can reduce text quality and introduce bias, making it easier for attackers to mimic watermarks through it. Public trust is crucial for both artificial intelligence generated text detectors and watermark algorithms. Imitation can damage public trust, make algorithms unusable, and hinder policy implementation.

As the same time, current researches overlooks the personalized watermarking, which means they are not able to inject watermark according to different users. In our method, however, by a personalized hashing function design, the watermark inject process is individual according to different user IDs. Each user will receive a specific watermark according to their ID, and this ensures that the specific user will be spotted when the watermark is retrieved.

\section{Overview}

\subsection{Threat Model}

The defender's goal is to embed watermark messages into the generated text in a way that is invisible to human readers while maintaining the model's normal behavior. We assume the defender has white-box access to the model, including its weights and generation strategies.
We assume the adversary's goal is to erase the injected watermark messages while preserving the unique language style of the content, particularly since these scenarios typically involve personalized LLMs. We also assume the adversary can only access the victim model in a black-box manner, querying the provided API through a limited number of user accounts. The adversary may sample texts generated by the victim model, analyze them to identify the watermarking rules, and edit the texts to remove the embedded watermarks. The adversary can modify the texts as long as they remain functional and retain their original language style. It is important to note that the adversary's ability to collect samples is limited due to user control systems in model protection scenarios, which restrict access to a large number of user accounts.

\subsection{Method Overview}

In the PersonaMark method, as shown in Figure \ref{fig:overview_mark}, for a given input prompt of a specific user who logs in, the model's generation strategy is first modified to sentence oriented beam search. For a given prompt, the model only generates the next sentence, marked with a sentence ending symbol such as comma, colon, exclamation and period. At the same time, we use diversity beam search, set a beam size, and directly sample multiple possible sentence sequences from the hidden space as candidate sentence sets. We uses a sentence structure extractor to extract the structural information of candidate sentence sets, and uses label sequence of dependency tree for representation. Due to the fact that the watermarked object in this article is a sentence structure, this sentence structure extractor has principle word-level robustness against synonym replacement attacks, as the sentence structure information of the text remains largely unchanged after the synonym replacement attack. Afterwards, we map and encode the sentence structure information with a hashing function from a hashing function database, in which the hashing functions are created when users log in. This hashing function distinguishes the candidate set into blue and green sentences with equal probability, represented by 0 and 1, respectively. After obtaining the mapping encoding values of the candidate sentence set, the final output sentence is only selected from the subset of candidate sentences with an encoding value 1, which makes the sentences output by the model mainly green. For general texts, based on hash function design, the number of blue-green sentences should be average. But for texts with our watermark, all the sentences generated will be denoted green. 

For the watermark detection process, the text paragraph to be detected will be firstly represented by the sentences' structure information by  a sentence structure extractor. Then the representations will be converted to encoding values with the hashing functions in the hashing database. Each encoding will be tested with a statistical Z-test. The existence or absence of this significant difference can be determined through probability statistics, ultimately determining the presence of watermark information in any given text. The specific user who generates this watermarked text will be determined since only the output of the hashing function which is used to generate this text will be positive for the Z-test detector. The user Id will be determined since the hashing function and the detection pipeline is user-specific.

\begin{figure}
    \centering
    \includegraphics[width=1\linewidth]{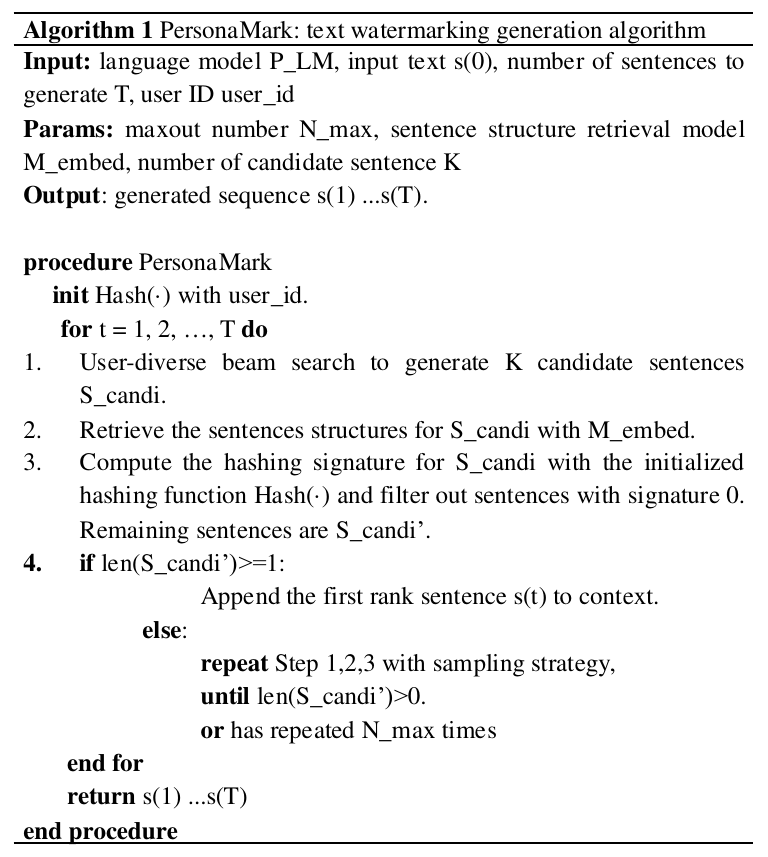}
    \label{fig:enter-label}
\end{figure}

\section{Method}

\begin{figure*}
    \centering
    \includegraphics[width=1\linewidth]{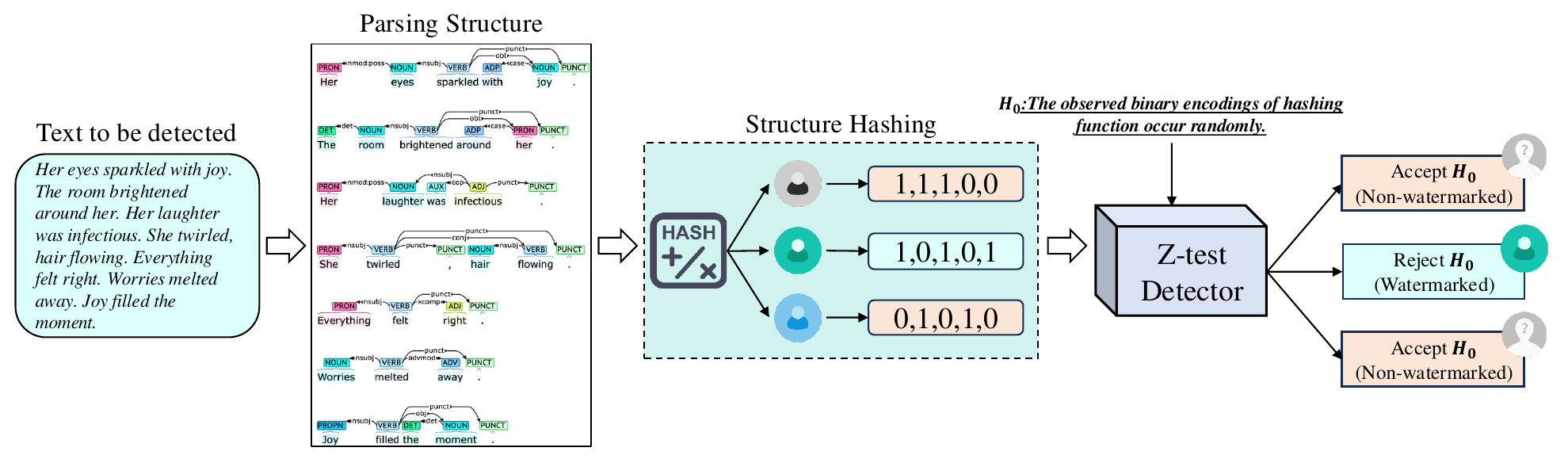}
    \caption{Overview of the personalized watermark detection process.}
    \label{fig:detect}
\end{figure*}

\subsection{Sentence structure representation}

We mainly use dependency syntax analysis to represent sentence structure information. The analysis results of dependency grammar help to deepen the understanding of sentence structure and meaning, thereby improving the execution efficiency of NLP tasks and providing a foundation for sentence structure encoding in this paper's task. The key advantage of dependency grammar lies in its cross linguistic applicability, especially for languages that have significant structural differences from inflectional languages such as English. The sentence structure extractor takes a sentence as input, analyzes the syntax structures, and generates a dependency tree. With each words in the sentence denoted with a dependency tag, the sentence structure extractor gives a string as output, which contains a sequence representing the corresponding sentence's structure. 

Extracting structure at the sentence level can better maintain the diversity of the text. The watermark information in this article is achieved through selective sampling of candidate sentences, without removing any words from the vocabulary, thus fully preserving the diversity at the lexical level. At the same time, the representation space of sentence structure is enormous. Even if some candidate structures are permanently removed, compared to word based approaches, the removed part of the sentences occupies a relatively small proportion of the overall potential sentence structure space. We set the structural threshold to 5, and adds watermarks to sentences with word lengths exceeding 5. If the sentence is short and the sentence structure space is small, rejecting some sentence structures may cause greater damage to the quality of the text. In terms of vocabulary size, if it is 65024 for a LLM, a segmentation of the vocabulary will make it reduce to 32512. This greatly reduces vocabulary diversity. Taking a mainstream dependency parsing system, spaCy\footnote{https://spacy.io/} as example, there are around 45 common dependency syntax labels based on dependency syntax analysis(`en\_core\_web\_sm'). With a threshold of 5 when not considering the punctuation mark "punch", there are ${45^{5}}$ candidate sets in the sentence structure space based on dependency syntax analysis. Although the actual sentence structure possibilities are constrained by linguistic rules and semantic plausibility, the range of selectable options remains substantial when considering Martin Cutts' recommendation that the average sentence length should ideally fall between 15 and 20 words\cite{cutts_what_2020}. The vast sentence structure space is much larger than the candidate word space, providing a solid foundation for imperceptible addition of watermark information and maintenance of text quality.

\subsection{Sentence-level generation}

In order to catch sentences as the processing objects, we improved the Beam Search generation strategy from multiple perspectives. Firstly, the ending strategy is task oriented optimized at multiple levels from the End of Sentence (EOS) perspective. Normal transformer-based LLMs are not able to stop at pre-defined punctuation during beam search process since they are trained under the special token `eos'. Our ending strategy allows the model to output only one sentence in a single computation, facilitating the addition of watermark information at the sentence level. Secondly, we use the diversity beam search to enhance the diversity of text generated content. In addition, in terms of beam scoring function, we achieve Shadow Fusion\footnote{Beam search: \url{https://opennmt.net/OpenNMT/translation/beam_search}} in the beam search stage from a post-processing perspective, which can improve the effectiveness of beam search and enhance the quality of text generated. Current LLMs are usually trained on large datasets to be base models and then supervised fine-tuned to be instruction following models, thus they all often a model set. After the EOS engineering, we selectively utilize this advantage and use the base models as matching auxiliary language models to adjust the scores of candidate beams, in order to improve the text quality.

To improve quality of output and to optimize the search process, hypothesis filtering is implemented, which dynamically eliminates candidate hypotheses that are unlikely to result in high-quality outputs. This method is particularly suitable for language models with a large number of possible outputs, as well as the framework that needs to generate a set of candidate sentences in this paper. The filtering rules include three dimensions: \textbf{L}, \textbf{R}, and \textbf{P}. \textbf{L: }Length limitation, which helps to avoid outputs that are longer or shorter than expected. \textbf{R: }Repetition punishment. If a hypothesis contains words that have already appeared in previous steps, its score can be reduced or filtered out directly. \textbf{P: }Prior knowledge. We filter out assumptions that violate common sense or pre-defined rules based on specific domain knowledge, which is implemented by a separate checker. The rule checker is independently completed at the end of the processing steps to achieve flexibility for potential changes. After Hypothesis Filtering, the obtained candidate sentence set will be encoded and filtered as watermark addition objects.

The hash encoding technology and sentence structure extractor mentioned in the previous sections are used to map the sentence structure of all sentences in the candidate sentence set into binary encodings. Sentences encoded as 1 remain unchanged, while sentences encoded as 0 are excluded from the candidate sentence set. This achieves the manipulation of the probability distribution of sentence structure encoding and ultimately adds the hidden watermark information. From the perspective of beam search optimization, this "0-1" mapping result filtering process can be seen as one of the components of hypothesis filtering.

\subsection{Signature-driven hashing and personalization}

Inspired by previous work\cite{kirchenbauer_watermark_2023,yang2023watermarking}, dividing the candidate space into two parts and manipulating the probability is an efficient way of injecting watermark. Previous work which manipulate the token probability usually divide the vocabulary into a green list and a red list and block the appearance of words in the red list. Vocabulary, or the word space or token space is pre-defined and limited, so it is easy to be divided using a dictionary. However, sentence space and the sentence structure space is not pre-defined. They are hard to listed and divided. We use a hashing method to project the potential sentence structure space into two parts instead of using a dictionary. 

The input of the hashing function is the output string of the sentence structure extractor. The input string is firstly hashed by a cryptography hash like BLAKE2 into a fixed length hash value. This value is later converted through a pipeline including hexadecimal transformation and integer transformation. The output of the hashing function is a binary encoding after a mod operation of 2.

Let s denote the a candidate sentence in the generating process at any step, and h(·) represents the string hash function like BLAKE2, and e(·) represents the sentence structure extractor. We utilize the output of the string hash function h(·) for generating a random binary value corresponding to s, as formalized as follows.
 \begin{equation}
b_{s} = \operatorname{RandomBinary}\left(h\left(e\left(s\right)\right)\right)
\end{equation}
Inheriting the features of BLAKE2 hashing algorithms, the simple hashing functions also have the advantages of nearly infinite input field, consistency of output for the same input, near-uniform nature, and evenly output domain. These features guarantee the successful division of the whole sentence structure space. The nearly infinite input field could handle the large amount of sentence structure that could exist. The consistency ensures that the watermark injection and watermark detection process could match with each other since the hashing output is exactly the same for the same input. As the output values guarantee uniform hashing with equal probability, they can be mapped to $[0, m-1]$, and the number on each value is also guaranteed to be almost equally likely if the value is modulo m. As we set m to 2, these advantages lead to that the mapped output is equally likely to be 0 or 1 for any given sentence structure string. Thus the sentence structure space is divided into two parts equally without a pre-defined dictionary. In other words, the text without watermark should have a roughly equal distrbution of sentences denoted 0 and 1. Based on this feature, watermark is injected by altering the distribution and highly raising the proportion of sentences denoted 1. After that, a statistical Z-test method could determine whether watermark signal exists. Such detection methods are commonly used in prior works\cite{kirchenbauer_watermark_2023,kirchenbauer2024reliabilitywatermarkslargelanguage,wu_dipmark_2023,hu_unbiased_2023} as well. Under the null hypothesis that the observed binary encodings in the text occur randomly, the expected proportion of encoding 1 will be 0.5. The Z-test detection can be formalized as follows, 
 \begin{equation}
 Z = \frac{\hat{p} - p_0}{\sqrt{\frac{p_0(1 - p_0)}{n}}}
\end{equation}
where ${p_0}$ is the expected proportion 0.5, $\hat{p}$ is the observed proportion of encoding 1 in the text, and n is the number of encodings which is the number of sentences in the detected text. By comparing the Z value with the values in Z-test table with a fixed significance $\alpha$, we can determine whether to reject the null hypothesis.

The normal injection process for the watermark is achieved without considering different users by now. And the watermark injection process is enhanced to fit personalized watermarking by making the hashing component special for different users. Our method's process relies on the hashing component to divide the sentence structure space into two parts, as mentioned above. We slightly enhance the hashing process to achieve a specialized hashing for different users, in order to achieve a personalized watermarking process for different users. Specifically, for a specific user, we concatenate the logged in user ID with the original sentence structure representation as the input of the hashing functions. Under this setting, the hashing process will be different according to the user IDs, even though the input sentence structure might be the same. Therefore, the hashing space will be different too, as each hashing function has its own corresponding user ID. We create a hashing function database, in which each user has his own hashing function with special hashing space. When it comes to watermark detection, only the matched hashing space will be positive to the hashing functions during different detection processes, since watermark injection and detection is a pair with the same hashing function. The text generated under hashing function A will not be denoted positive by a detection process under hashing function B. When a hashing function matches the text to be detected, the corresponding user ID will also be determined.

\section{Experimental Setup}

\subsection{Evaluation Metrics}
We selects perplexity, aspect based text sentiment analysis, key information consistency, and reading ease as indicators to measure the impact and intervention of watermarking methods on model behaviors. The perplexity is usually used for estimating text quality, and it is calculated using formula below.
\begin{equation}
\begin{split}
\operatorname{PPL}(T) &= P(t_{1}, t_{2}, \ldots, t_{N})^{-\frac{1}{N}} \\
&= \exp \left(-\frac{1}{N} \sum_{i=1}^{N} \log P(t_{i} \mid t_{1}, t_{2}, \ldots, t_{i-1})\right)
\end{split}
\end{equation}

It can be calculated by the Nth root of the reciprocal of the joint probability of the sentences in the test set, where N is the average number of words in the sentence. Where T represents a set of text sequences, including tokens such as t1, t2 to tN, and the probability in the perplexity formula can be calculated by a language model. The aspect based sentiment polarity is calculated using the deberta-v3-base-absa-v1.1 model\footnote{deberta\-v3\-base\-absa\-v1.1:\url{https://huggingface.co/yangheng/deberta-v3-base-absa-v1.1}}. We utilize aspect based sentiment analysis rather than normal sentiment analysis since the prompts in the experiment are open-end generation task. Normal sentiment analysis tends to have greater fluctuations. We set the subjects of sentences as aspects and calculate the sentiment score according to the subjects. In this way, the average sentiment polarity of the model is determined as a metrics indicating the behavior of the models. The key information consistency is represented by the similarity obtained from the Sentience Transformer (all-MiniLM-L6-v2) model\footnote{Sentence Transformers documentation:\url{https://www.sbert.net/docs/sentence_transformer/pretrained_models.html}} after summarization using the facebook/bart-large-cnn model\footnote{facebook/bart-large-cnn:\url{https://huggingface.co/facebook/bart-large-cnn}}. For a given prompt, our approach's aim is to let the models generate the same key components as the models without watermark, though the responded sentences might be different. In addition, the reading ease indicator is used to assist in evaluating the behavioral changes of models when generating text. The reading ease score is based on the Flesch Reading Ease Score (FRES)\footnote{Flesch Reading Ease and Flesch Kincaid Grade Level:\url{https://readable.com/readability/flesch-reading-ease-flesch-kincaid-grade-level/}}, which is used to evaluate the readability of an article or sentence from the perspective of the age required to read the sentence. For the English language, the formula for calculating sentence difficulty is as follows:
\begin{equation}
R.E.=206.835-0.846*wl-1.015*sl
\end{equation}

where R.E. refers to readability score, wl refers to the number of syllables per 100 words, sl corresponds to the average word count in each sentence, and the total R.E. score ranges from 0 to 100. The higher the score, the easier it is to read.

\begin{table*}[!t]
\centering
\caption{Evaluation results of different model settings. $\Delta$ represents the absolute value of the difference between the watermarked text and the text without watermark. $\uparrow$ indicates higher values are preferred. $\downarrow$ indicates lower values are preferred.  Compared with the baseline KGW, the text generated by PersonaMark is closer to the text of the original model across several dimensions, indicating less perturbation to the model behavior.}
\resizebox{\textwidth}{!}{
\begin{tabular}{llcccc}
\toprule
\textbf{Models} &
  \textbf{Method} &
  \multicolumn{1}{l}{\textbf{Perplexity$\downarrow$}} &
  \multicolumn{1}{l}{\textbf{Sentiment Polarity($\Delta\downarrow$)}} &
  \multicolumn{1}{l}{\textbf{Alignment$\uparrow$}} &
  \multicolumn{1}{l}{\textbf{Readability($\Delta\downarrow$)}} \\
  \midrule
\multirow{4}{*}{Gemma-2B-IT}              & Without Watermark & 8.83          & 0.63               & -             & 58.70               \\
                                          & KGW               & 12.32         & 0.66/0.03          & 0.56          & 46.91/11.79         \\
                                          & KGW-Persona       & 6.53         & 0.68/0.05             & 0.58      & 55.52/3.18             \\
                                          & \textbf{PersonaMark(Ours)}   & \textbf{2.74} & \textbf{0.62/0.01} & \textbf{0.63} & \textbf{55.53/3.17} \\
\midrule                                          
\multirow{4}{*}{Mistral-7B-Instruct-V0.3} & Without Watermark & 1.78          & 0.93               & -             & 53.49               \\
                                          & KGW               & 3.81          & \textbf{0.95/0.02} & 0.60          & 44.09/9.40               \\
                                          & KGW-Persona       & 4.39          & 0.83/0.1            & 0.58       & 41.68/11.81              \\
                                          & \textbf{PersonaMark(Ours)}   & \textbf{2.10} & 0.83/0.1           & \textbf{0.64} & \textbf{44.56/8.93}      \\
\midrule
\multirow{4}{*}{OLMo-7B-Instruct}         & Without Watermark & 2.37          & 0.71               & -             & 59.65      \\
                                          & KGW               & 7.54          & 0.91/0.2               & 0.59          & 43.37/16.28               \\
                                          & KGW-Persona       & 6.53         & 0.86/0.15           & 0.56        & 48.81/10.84            \\
                                          & \textbf{PersonaMark(Ours)}   & \textbf{2.39}         & \textbf{0.70/0.01}               & \textbf{0.64} & \textbf{49.36/10.29}               \\
\midrule
\multirow{4}{*}{Phi-3.5-mini-Instruct}                    & Without Watermark & 2.10 & 0.82               & -             & 44.92      \\
                                          & KGW               & 6.84          & 0.92/0.1               & 0.10          & 38.72/6.20               \\
                                          & KGW-Persona       & 8.05        & \textbf{0.73/0.09}           & 0.11     & 37.07/7.85               \\
                                          & \textbf{PersonaMark(Ours)}   & \textbf{3.19}          & 0.92/0.1      & \textbf{0.59} & \textbf{39.99/4.93}   \\           
\bottomrule
\end{tabular}
}

\label{main}
\end{table*}

\subsection{Baselines}
We select the most commonly used watermarking method KGW as the baseline method. The KGW method is based on manipulating the probability of logits, dividing the vocabulary into green and red word lists with randomness. The logits corresponding to the red word list are adjusted to 0 or a penalty score is applied to make them appear less often. Finally, statistical tests are used to determine whether any given text has this specificity. Due to its concise design and model adaptability, it has become a commonly used baseline in the field of text watermarking, and lots of current research on text watermarking work on it for improvement. We measure the interference of watermarking methods on model generation behavior through a comprehensive evaluation of generated text, and combines text correctness indicator to control the degree of synonym attacks within a reasonable range. The text correctness is calculated by the following formula where C is the number of errors computed with LanguageTool\footnote{LanguageTool, style and grammar checker for 25+ languages:\url{https://github.com/languagetool-org/languagetool}}.  \begin{equation}
\text{TextCorrectness} = 
\begin{cases}
\log_{2}\left(\frac{1}{c}\right), & \text{if } C \neq 0 \\
1, & \text{if } C = 0
\end{cases}
\end{equation}When the proportion of synonym attacks reaches 40\%, through manual evaluation and GPT-4 based quality evaluation, it is found that the text already contains too many syntactic, lexical, and other errors. Therefore, the proportion of synonym is limited to 40\% to fit real-world scenarios. We select four open-source model series as the base model for evaluation including Gemma-2B, Mistral-7B, OLMo-7B and Phi-3.5-mini. In order to conduct experimental comparisons under resource constraints, the random sampling strategy is used as an auxiliary supplementary method. When a smaller beam width causes the model to fall into a local minimum, disturbance intervention is performed by increasing randomness.

Current text watermarking methods lacks the ability to inject different watermark for different users. To facilitate comparison, we modified the KGW method to KGW-Persona to forcefully let it imitate PersonaMark's strategy, making it suitable as a baseline for a personalized setting. The gamma parameter of KGW is set to 0.5 for dividing the vocabulary into two equal parts. The KGW divides the vocabulary with randomness initialized by previous tokens into green list and red list and calculates the proportion of green list words during watermark detection. With randomness, each word in the vocabulary will be able to appear in text.  We froze the green list for specific users. Each user has a fixed green list and red list division with a frozen green-red division line. For each user the green list will be different. In this way we define different watermark injection process for different users.

\subsection{Platform and Generation Settings}
We conducted our experiments on a system equipped with an Intel(R) Xeon(R) Gold 6330 CPU @ 2.00GHz processor, and an NVIDIA A100-SXM4-80GB GPU, complemented by 1511G of memory. The parameter settings for our experiments are as follows:
The 'temperature' parameter, which controls the diversity of the auxiliary stochastic sampling strategy, was set to 2.0. Max\_length was set to 32, indicating the maximum token length for the summarization model. Min\_length was set to 10, representing the minimum token length for the summarization model. The threshold for statistical Z-test was set at 4. Top\_k was set to 50, defining the number of words in the sampling space for the stochastic sampling strategy. Top\_p was set to 0.95, indicating the cumulative probability value for the sampling space in the stochastic sampling strategy. Sent\_num was set to 16, which is the number of sentences generated by the PersonaMark method. Num\_beam\_groups was set to 12, representing the number of groups used in the Diverse Beam Search. Diversity\_penalty was set to 1.5, a parameter that encourages diverse text generation. Repetition\_penalty was set to 1.1, a parameter that reduces the generation of repetitive content. Length\_penalty was set to 0.7, a parameter that encourages the generation of shorter sentences. Num\_beams was set to 24, indicating the number of beams used in a single search for Beam Search.

\begin{figure}[!t]
    \centering
    \includegraphics[width=1\linewidth]{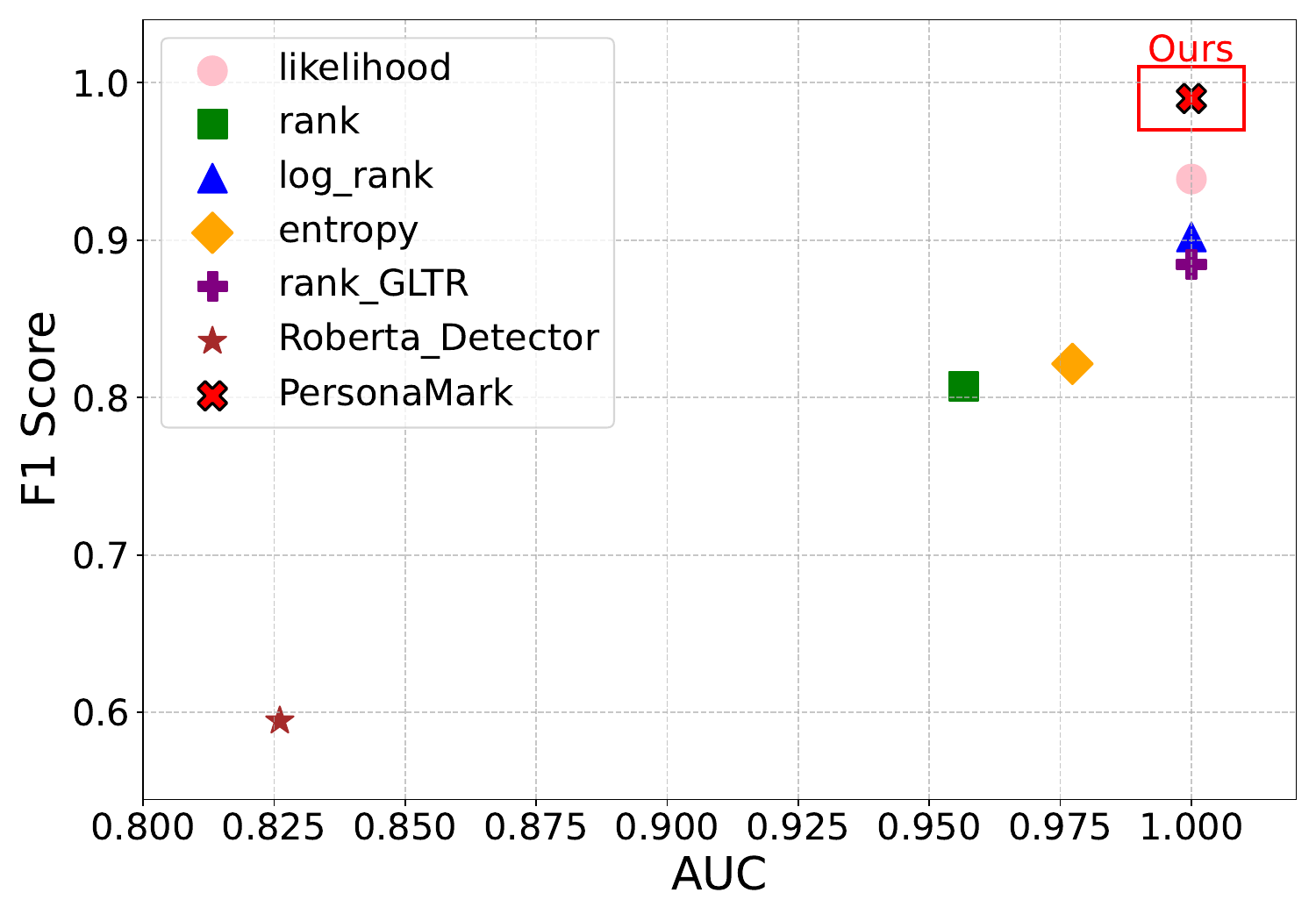}
    \caption{AI-generated text detection performance on content generated by the PersonaMark. As watermarking techniques become one of the approaches for AI-generated text detection, we compare PersonaMark with five metric-based methods and a model-based method of this task\cite{mgtbench}. Results show that our method aiming for multi-user attribution still retains the ability of zero-bit watermark on AI-generated text detection , with F1 value and AUC value close to the maximum value.}
    \label{fig:point}
\end{figure}

\begin{figure*}[!t]
    \centering
    \includegraphics[width=1\linewidth]{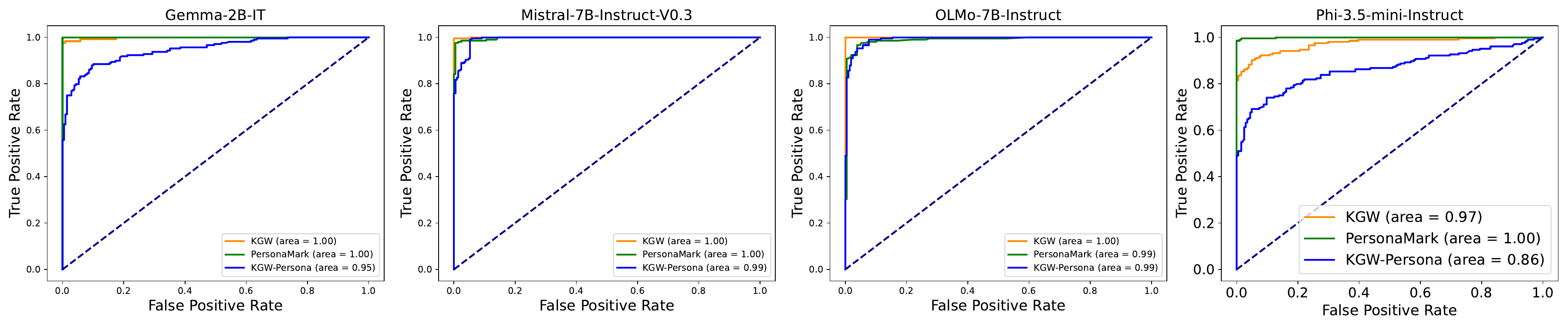}
    \caption{Watermark detection performace on four models measured by AUC curve and score. The green, orange, and blue lines denote performance of our PersonaMark method, KGW, and KGW-Persona. The titles are different LLMs' name. Our PersonaMark achieves high AUC across different models. KGW‘s AUC, when converted to KGW-Persona to fit in the peronalized setting, drops much. }
    \label{fig:roc}
\end{figure*}

\begin{figure*}[!t]
    \centering
    \includegraphics[width=1\linewidth]{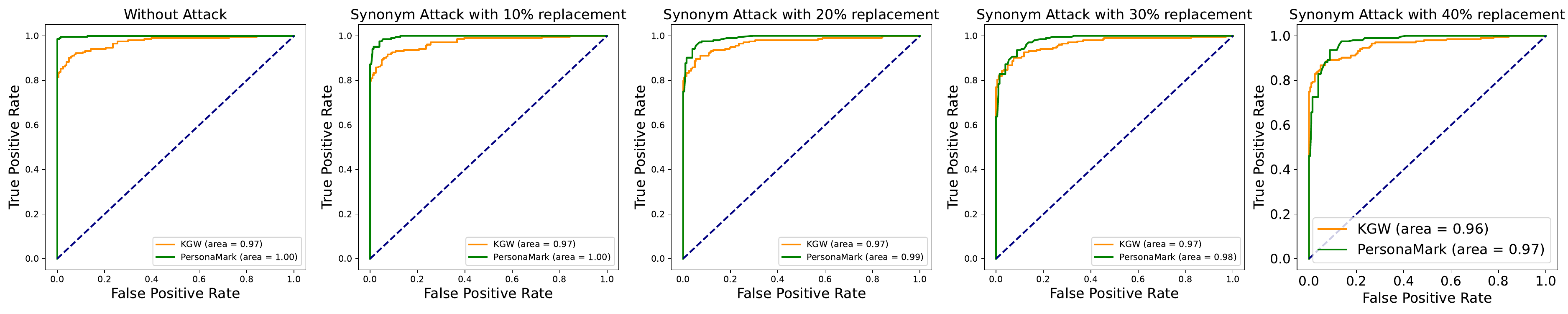}
    \caption{Watermark performance of texts genereated by Phi-3.5 model. The titles denotes the attacking strength defined by word replacement probability. The green line and orange line denote performace of our PersonaMark and KGW respectively. After the synonym attacking, the AUC score drops with the attacking strength while our PersonaMark shows better robustness than KGW.  }
    \label{fig:roc_attack}
\end{figure*}

\section{Results}
\subsection{Effectiveness Analysis}
\begin{figure*}[!ht]
    \centering
    \includegraphics[width=1\linewidth]{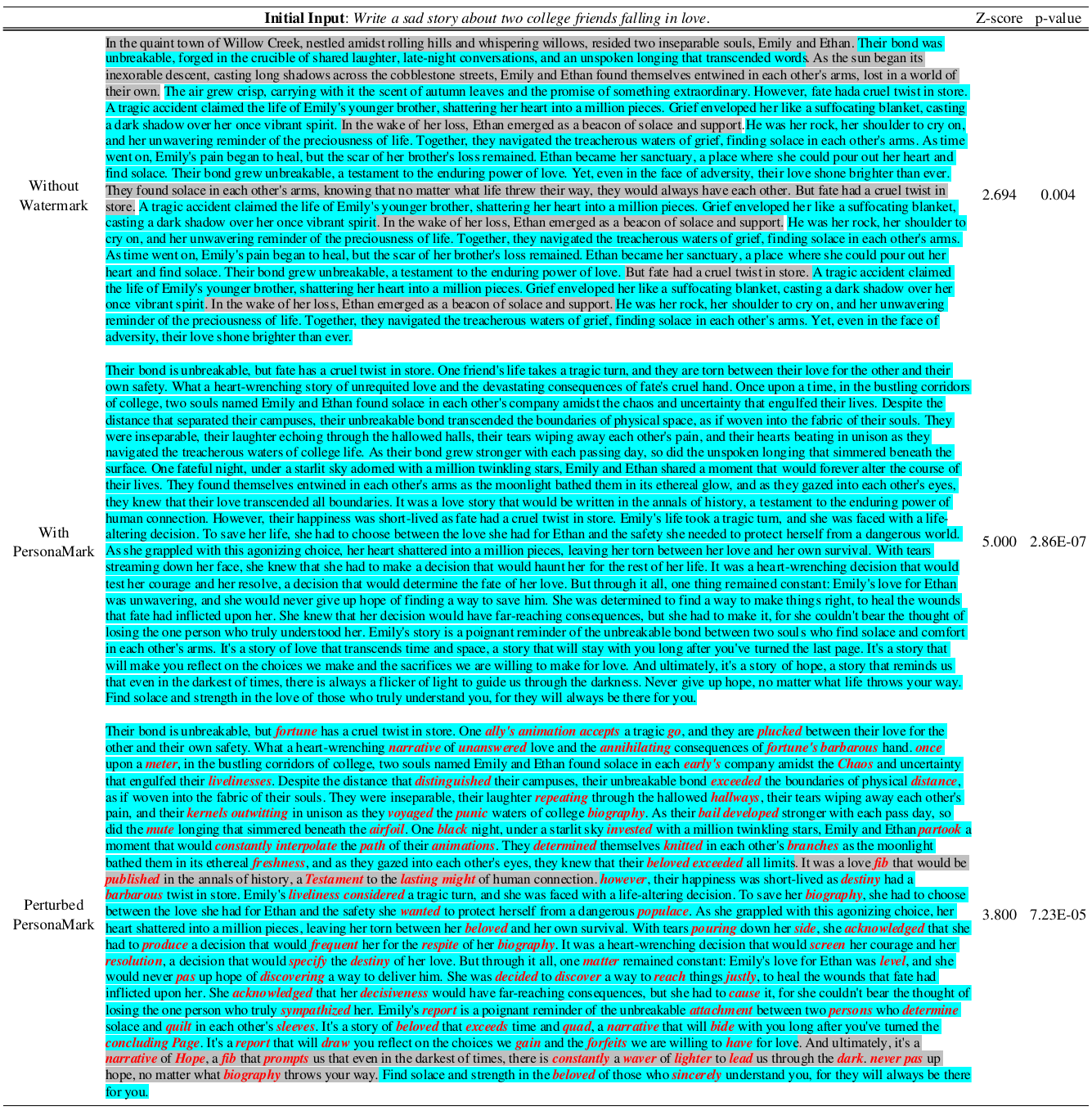}
    \caption{
A case study with PersonaMark. Let the same open-end generation prompt be the initial input. The text generated by Gemma without watermark contains both bit-0 sentences and bit-1 sentences, denoted as grey and light blue respectively, the number of which should be roughly equal. Its Z-score is low and p-value is high. The null hypothesis will not be rejected and this piece of text will not be detected as watermarked. The text generated by Gemma applied PersonaMark algorithm contains only bit-1 sentences, denoted light blue. The Z-value is high and p-value is extremely low, which is enough to reject the null hypothesis. \textbf{\textit{Red}} denotes the word generated by the word-substitution attack method. The text of PersonaMark is perturbed by large proportion of synonym attack to be Perturbed PersonaMark text. Only three bit-1 sentences is hashed as bit-0 after the synonym attacking perturbation, thus the Z-score and p-value almost remains and we can reject the null hypothesis with a high confidence(low p-value).}
    \label{fig:case}
\end{figure*}

\textbf{Watermark Strength.} As shown in Table \ref{main}, Figure \ref{fig:roc}, and Figure \ref{fig:roc_attack}, 
KGW's watermark detection performance on Phi-3.5 model is lower than other models. And when KGW is converted to KGW-Persona with a fixed green list to fit personalized settings, the AUC score drops significantly. This performance decline can likely be attributed to the vocabulary size: Phi-3.5 supports a vocabulary of up to 32,064 tokens, whereas other models, such as Gemma-2B-it, use a much larger vocabulary of 256,128 tokens. KGW and KGW-based methods inject their watermark by altering the token probability distribution. A restricted vocabulary size restrains the potential watermarking space, thus leading to worse performance and lower robustness. This finding supports the feasibility of PersonaMark, as its sentence-level watermarking operates in a far larger space than the fixed token vocabulary of LLMs. The observed performance drop is unlikely to be due to the number of parameters or model size, since Gemma-2B, which is smaller than Phi-3.5 (3.8B parameters), performs worse.

\textbf{Model Behaviors.} We measure model behavior with metrics describing features of the text outputs, including perplexity, ABSA sentiment polarity, reading ease score, and key information consistency. When compared to KGW, the mainstream text watermarking baseline, PersonaMark achieved the best results in terms of perplexity, indicating minimal interference with the text generation process. This comes from the level of watermarking, as KGW-based method manipulate the probability distribution of words. The probability of distribution is one of the goal of pretraining, and its modification introduces much bias. However, PersonaMark manipulates the sentence-level distribution, which is less intrusive because it operates over a larger space, and does not alter the internal distribution of tokens within sentences.

In terms of ABSA average sentiment polarity, semantic key information consistency, and sentence readability, PersonaMark has the smallest difference compared to the text generated by models without watermarking. The reading ease score for baselines drops more than that for PersonaMark. This may come from their watermarking scheme that part of vocabulary is randomly blocked as red list while common words are not evenly distributed in the vocabulary, which reduces the appearing probability of some common words and increases the appearing probability of some uncommon words. The increase of baselines in the frequency of uncommon words increases the difficulty of reading, while PersonaMark preserves more common words. In terms of key information consistency especially, the similarity values of PersonaMark is much better than the baseline method with the same base model and input instructions. This is not only comes from the level of watermarking, but also comes from the manipulation frequency. KGW will manipulate each word's probability in a sentence while the manipulation frequency of PersonaMark is lower as each sentence is manipulated once. What's more, the i-th word's green list and red list division depends on previous words in the baselines, so the bias of introducing watermark is reinforced and accumulated sequentially, making the model further away from the original state.

\begin{table}[!t]
\centering
\caption{Win rate of our PersonaMark on Fluency and Content Consistency}
\resizebox{0.49\textwidth}{!}{
\begin{tabular}{lcc}
\toprule
\textbf{Models} & \textbf{Fluency (\%)} & \textbf{Content Consistency (\%)} \\
\midrule
Gemma-2B-IT & 80 & 60 \\
Mistral-7B-Instruct-V0.3 & 100 & 85 \\
OLMo-7B-Instruct & 75 & 65 \\
Phi-3.5-mini-Instruct & 100 & 100 \\
\bottomrule
\end{tabular}
}
\label{tab:fluency_consistency}
\end{table}

\textbf{Human Evaluation.} We conduct a human evaluation by sampling 20 items from each model's outputs, involving four evaluators comprising two doctoral students and two master students. The evaluation focuses on two dimensions: Fluency and content consistency. The evaluators are provided with the prompt used for generation, the original model output without watermark, the KGW-Persona output, and the PersonaMark output.. They are instructed to select the superior watermarked text based on its fluency and its consistency with the original output. The former dimension aims to assess the quality of texts, while the latter aims to indicate changes in model behavior when using identical prompts. The findings presented in Table \ref{tab:fluency_consistency} indicate that PersonaMark achieves significantly higher average win rates in both fluency and content consistency, with respective values of 88.75 and 77.75. These results markedly surpass those of the multi-user baseline, KGW-Persona. Notably, the Phi-3.5 model entirely fails to respond to the original prompt under KGW-Persona settings. For instance, it generates irrelevant content such as math, code, or biological knowledge for story generation prompts. In contrast, under our PersonaMark method, the model retains the ability to follow instructions promptly. The underlying cause for this discrepancy may be analogous to the rationale previously elucidated regarding watermark strength. The Phi-3.5 model exhibits a comparatively limited vocabulary size relative to other models, resulting in a residual vocabulary of only 16,032 tokens after deployment of the KGW-Persona watermarking. This substantial reduction in vocabulary severely impairs the model's generative capabilities. This limitation underscores the inherent performance degradation of the KGW-like methods, attributable to the incongruity between logits manipulation and the training and inference processes of large language models.

\textbf{Watermark Recognition Performance.} In terms of watermark recognition performance, the F1 and AUC values of our method are better than the baseline, with better classification performance and close to the theoretical maximum value. Rather than a non-interpretable deep learning detector, the Z-test of null and alternative hypotheses has statistical interpretability while improving the recognition performance of the detector. In terms of robustness, PersonaMark demonstrates high robustness against synonym attacks. This is due to the principle advantage of sentence structure representation strategy. When synonym attacks are limited by grammar and quality rules, they will not change the sentence structure representation, making it impossible to effectively attack our method.

\begin{table}[t]
\centering
\caption{Detection time of different user amounts.}
\begin{tabular}{cc}
\toprule
User amount & Runtime (second) \\
\midrule
1 & 0.003141 \\
1000 & 0.039041 \\
10000 & 0.386027 \\
100000 & 3.857007 \\
\bottomrule
\end{tabular}
\label{tab:runtime}
\end{table}

\textbf{Watermark Detection Speed.} During the watermark detection process, since the text might be generated with one of the hashing functions in the database. In order to determine exactly which hashing function matches the text or no one matches the text(means it is not watermarked by our approach), each user's hashing function need to operate the hashing one time. For 100 users, the hashing process needs to repeat 100 times. When the number of users exceeds 1000, the detection speed increases linearly with the number of users. However, the detection speed is also fast due to the fast computation speed of hashing algorithms. As show in Table \ref{tab:runtime} it takes 3.86 seconds when there are 100,000 users. The detection speed could be faster with more CPU cores or with parallel computation.

\textbf{AI-generated text Detection.} We also compare PersonaMark with several AI-generated text detectors to explore its ability under AI-generated text detection scenarios. Results show that our approach retains the ability for AI-generated detection task. In terms of the effectiveness of text detection generated by artificial intelligence, the F1 score of the detection results has significantly improved compared to the seven commonly used detection methods.

\subsection{Case Study}
As shown in Figure \ref{fig:case}, the text generated by
Gemma without watermark contains both bit-0 sentences and bit-1 sentences, denoted as grey and light blue respectively, the
number of which should be roughly equal. Its Z-score is low and p-value is high. The null hypothesis will not be rejected and
this piece of text will not be detected as watermarked. The text generated by Gemma applied PersonaMark algorithm contains
only bit-1 sentences, denoted light blue. The Z-value is high and p-value is extremely low, which is enough to reject the null
hypothesis. The text of PersonaMark is perturbed by large proportion of synonym attack to be Perturbed PersonaMark text. Only
three bit-1 sentences is hashed as bit-0 after the synonym attacking perturbation, thus the Z-score and p-value almost remains and
we can reject the null hypothesis with a high confidence(low p-value). The P value refers to the probability of a more extreme result than the sample observation when the null hypothesis is true. If the P value is very small, it means that the probability of the null hypothesis is very small. If it occurs, according to the principle of small probability, we have reason to reject the null hypothesis. The smaller the P value, the more reason we have to reject the null hypothesis.

According to Figure \ref{fig:case} in the Appendix, case texts shows that without adding watermark, the proportion of blue and green sentences in the text generated by the model is roughly equal. After using the PersonaMark method to add watermarks, the model generates only green sentences. Under the synonym substitution attack, the bold and skewed words are replaced after the attack. It shows that the proportion of green sentences changing to blue sentences is extremely low, maintaining the integrity of the watermark information. We assume the reason for why synonym substitution attacks lacks efficiency on PersonaMark is that the replaced words and the original words usually have the same part of speech and functionality in the sentences, thus the sentences structures remain the same. The synonym substitution attacking changes the structures as long as it considers the grammar and usability of texts. 

After the synonym substitution attack, a large proportion of text generated under PersonaMark is replaced with similar words, such as fate replaced as fortune. However, most of the sentences’ encoding is unchanged, which is denoted by color. In the second sentence, the word “takes” is replaced with “accepts”, but the sentence structure remains the same. When synonym substitution attack is restrained by grammar rules like “keeping the tense‘, the sentence structure is likely to remain. In the first grey sentence in the perturbed PersonaMark text, the synonym’s meaning leads the sentence to another meaning, while the structure extractor falsely classified the phrase ‘fib that’. In the case of the second grey sentence in perturbed text, all other words’ tags remain the same, while “guide” is replaced with “lead”. This is an classification error of structure extractor. In the last grey sentence, the replaced words’ functions and tags remain while the tag of the word “what” changes due to its following word. We assume that deep learning based sentence structure extractor is more sensible to word replacement and the more detailed dependency tag system will be more vulnerable. However, current framework is resilient enough for synonym substitution attacks, whose p-value is extremely low after perturbing with enough confidence for detection.

\section{Discussions}

\subsection{Limitations}
Since the watermark injection process leverages beam search, our method necessitates more computing resources and inference time compared to non-watermarking generation processes. While the method's high scalability ensures that increased user numbers do not incur additional costs, the inference time for individual users is affected by the watermarking injection. Given that our method relies on sentence structures, further investigation might be required in contexts where sentence structures are less significant and do not convey stylistic features. Additionally, because PersonaMark operates at the sentence level, it is not applicable when users generate only a single sentence, representing a potential limitation. Nevertheless, in typical use cases, models are generally requested to produce outputs at the paragraph level or longer, which include multiple sentences. Moreover, even if one were to request single sentences repeatedly, the watermark would still be injected, thereby maintaining the method's effectiveness.

\subsection{Ethical Considerations}
Our research on PersonaMark has been conducted with careful consideration of potential ethical implications across multiple dimensions.

\textbf{Data Privacy}. PersonaMark utilizes a personalized hashing function to generate unique watermarks for each user. To achieve this, we ensure that no personally identifiable information (PII) is collected, stored, or transmitted. The algorithms could be achieved with anonymous IDs without any user data.

\textbf{Transparency}. We have prioritized transparency by providing detailed documentation on how PersonaMark operates, including the methodologies for watermark embedding and detection. By making the watermarking process understandable, we empower users to make informed decisions about its implementation.

\textbf{Minimizing Impact on Model Performance}. A critical ethical consideration is ensuring that the introduction of watermarking does not degrade the performance of LLMs. Through extensive evaluations, we have demonstrated that PersonaMark maintains high text quality, preserves the natural behavior of the models. By optimizing the watermarking process to be subtle and minimally invasive, we mitigate potential negative impacts on user experience and model functionality.

\textbf{Compliance with Legal and Ethical Standards}. Our research complies with relevant data protection regulations, including the General Data Protection Regulation (GDPR) and the California Consumer Privacy Act (CCPA). We have conducted thorough reviews to ensure that PersonaMark adheres to legal standards concerning data privacy and user consent. Additionally, we follow ethical disclosure practices by transparently reporting our methodologies, findings, and any limitations associated with PersonaMark. This openness facilitates accountability and fosters trust within the research community and among end-users.

\section{Conclusion}

In this paper, we presented PersonaMark, a novel personalized text watermarking scheme designed to address critical issues of copyright protection, accountability, and traceability in the context of customized LLMs. We employ sentence structure information as the medium for watermark injection. First, sentence structure representation is derived from syntactic analysis and we encode structural information using a carefully designed hash function. Next, utilizing a sentence-level beam search strategy, we sampled candidate sentences with encoding bit-1 information and successfully integrated the watermark into the generated text seamlessly. At the same time, by adding the user ID to the input of the hash function, we construct a hashing function database corresponding to specific users. Personalized hashing function provides personalized watermarking injection process. We achieve personalized watermarking by adding different watermarks for different users, which means the generated text under any user's account will contain his unique user ID. Though watermark detection based on the hashing function database, the user ID could be retrieved.

We propose new evaluation indicators related to model behavior for assisting the development of text watermarking techniques, and design experiments to measure the intervention and impact of watermarking algorithms on model behavior from a textual perspective, verifying the effectiveness of the algorithm proposed. The experimental results show that under comprehensive indicators PersonaMark has the least interference and impact on the model generation process, making it the first text watermarking method that achieve personalized watermarking without destroying model performance.

\bibliographystyle{plain}
\bibliography{references}

\begin{thebibliography}{10}
\bibitem{noauthor_new_nodate_criminals}
New report finds that criminals leverage {AI} for malicious use – and it’s not just deep fakes.
\end{thebibliography}
\end{document}